# Orbital Angular Momentum (OAM) Vertical-Cavity Surface-Emitting Lasers

**Huanlu Li[1, 2], David Phillips[3, 4], Xuyang Wang[1], Daniel Ho[1], Lifeng Chen[1], Xiaoqi Zhou[3], Jiangbo Zhu[5], Siyuan Yu[1, 5], Xinlun Cai[5]**

[1] Department of Electrical and Electronic Engineering, University of Bristol, University Walk, Bristol, BS8 1TR.

[2] School of Engineering, Rankine Building, Oakfield Avenue, University of Glasgow, Glasgow G12 8LP, UK.

[3] H.H. Wills Physics Laboratory, University of Bristol, Tyndall Avenue, Bristol BS8 1TL, UK.

[4] Department of Physics and Astronomy, University of Glasgow, Glasgow G12 8QQ, UK.

[5] State Key Laboratory of Optoelectronic Materials and Technologies and School of Physics and Engineering, Sun Yat-sen University, Guangzhou 510275, China.

E-mail: caixlun5@mail.sysu.edu.cn, david.phillips@glasgow.ac.uk

**Abstract**

Harnessing the Orbital Angular Momentum (OAM) of light is an appealing approach to developing photonic technologies for future applications in optical communications and high-dimensional Quantum Key Distributions (QKD). An outstanding challenge to the widespread uptake of the OAM resource is its efficient generation. We design a new device which can directly emit an OAM-carrying light beam. By fabricating micro-scale Spiral Phase Plates (SPPs) within the aperture of a Vertical-Cavity Surface-Emitting Laser (VCSELs), the linearly polarized Gaussian beam emitted by the VCSEL is converted into a beam carrying specific OAM modes and their superposition states with high efficiency and high beam quality. The innovative OAM emitter opens a new horizon in the field of OAM-based optical and quantum communications, especially for short reach data interconnects and Quantum Key Distribution (QKD).

**Introduction**

Optical beams with phase singularities, also known as optical vortices, were firstly discussed by Nye and Berry in 1974[1], who identified singularities within randomly scattered fields. In 1992, Allen *et al.* recognized that Orbital Angular Momentum (OAM) is a natural characteristic of all helical phased light beams, and could be readily generated in a standard optics lab[2]. Beams carrying OAM are attractive as they offer a theoretically unbounded number of possible states, and have therefore been a subject of great interest for a variety of fundamental and applied research activities at both the classical and single photon level, including communication[3], optical manipulation[4], optical microscopy[5], remote sensing[6] and quantum information[7]. Beams with a well-defined state of OAM have a complex field characterized by $\exp(il\varphi)$, where $\varphi$ is the azimuthal around the optical axis, and $l$ is the topological charge, describing how many integer multiples of $2\pi$ the phase changes by in one circuit of the vortex core. They are created by imposing an azimuthally dependent phase structure onto the beam. This is routinely achieved by using bulk optical components such as Spatial Light Modulators (SLMs). Despite offering a great flexibility, SLM suffers a slow response time limited to milliseconds, and have no clear route to miniaturization. In emerging OAM-based classical or quantum information systems both either on optical fibers[8] or free space[9] transmission, a key requirement is integrated components that generate OAM-carrying beams in a compact, cost-effective and reconfiguration way.

Recently, several silicon photonics integrated OAM emitters that convert planar waveguide modes into free-space OAM modes[10–12] have been reported as a potential candidate for future communication systems. These are significantly more compact and robust than their bulk optics counterparts. However, as passive devices, they require external lasers sources, which not only increases the cost and the complexity but also reduces the power efficiency that is critical for applications. Therefore, power-efficient active OAM-emitting devices that can be integrated in large scale and fabricated in large quantities at low cost are highly desirable.

Here, we report an innovative approach to generating specific OAM-carrying beams by integrating micro-sized Spiral Phase Plates (SPPs) in the aperture of Vertical-Cavity Surface-Emitting Laser (VCSELs). The VCSEL device represents an attractive light source in many industrial applications due to its low fabrication cost, high energy-efficiency, circular beam profile and high intrinsic modulation bandwidths[13,14]. VCSEL-like devices have also been intensely researched as single photon sources[15] and as low-cost sources in Quantum Key Distribution (QKD) systems[16]. By integrating VCSELs with mass-manufacturable structures, high purity OAM modes and their superposition states could be generated while maintaining advantages in cost and power efficiency. Photonic components can be produced not only highly suitable for OAM based optical communications systems, but also for quantum systems[17,18].

**VCSEL Devices and Fabrication of SPPs**

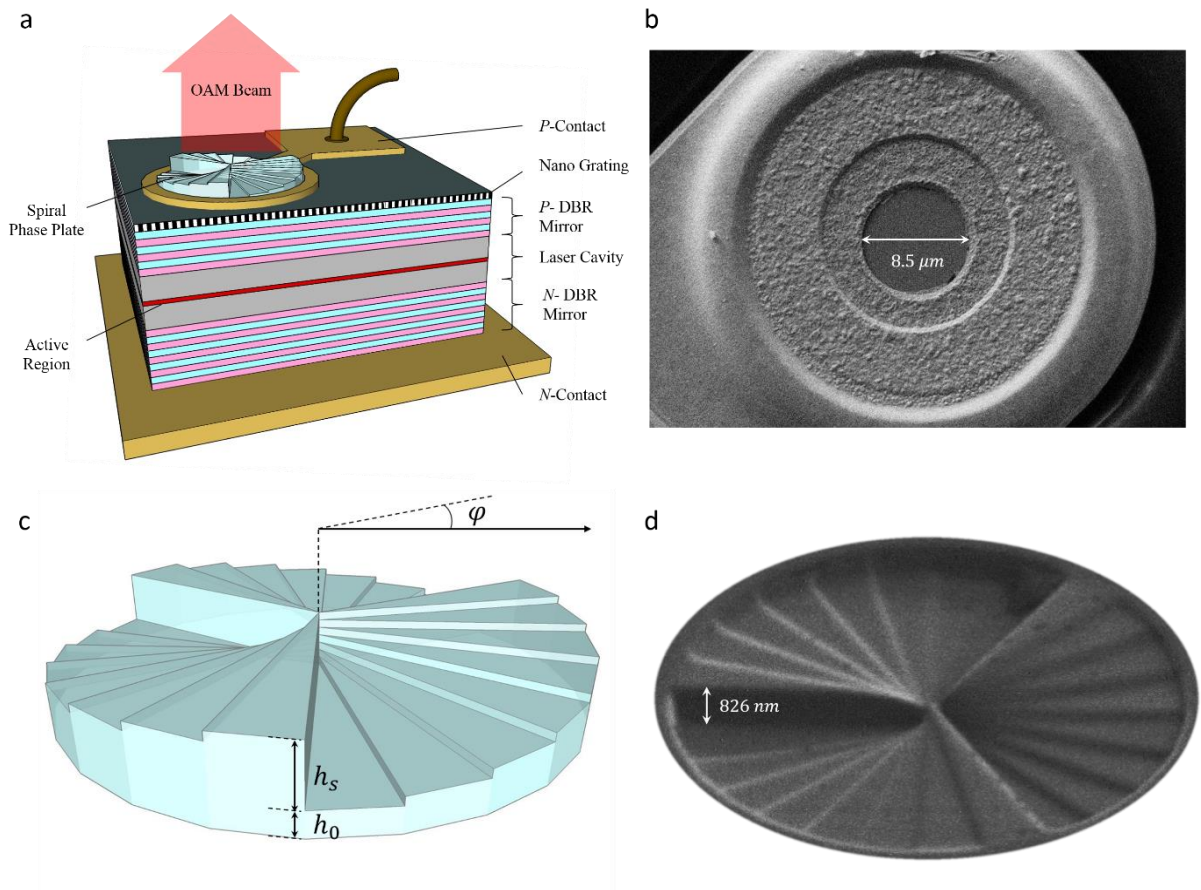

Figure 1. Structure of the VCSEL device with integrated SPP. (a) Schematic of the VCSEL device with an integrated SPP. (b) The SEM image of the VCSEL device before SPP fabrication. (c) Sketch of a stepped SPP with topological charge of +3. (d) The SEM image of the fabricated stepped SPP.

Fig. 1a illustrates the structure of a commercially available single mode and linear polarized VCSEL. The laser cavity consists of a lower distributed Bragg reflector (DBR) mirror, a laser active layer and an upper DBR mirror. The devices include a nano-grating in their aperture surface for polarization control so that linearly polarised beams are emitted. When the electric current is injected through the top P-contact and the bottom N-contact, a Gaussian beam is emitted from the aperture at a wavelength of 860nm with a typical threshold current of 1mA. A maximum output power of 4 mW can be achieved with an injected current of 6 mA. Fig. 1b shows the SEM image of the aperture of the VCSEL with a size of 8.5 $\mu m$ in diameter.

An $l$-fold helical phase front could be imprinted by way of introducing a SPP in the aperture of the VCSEL. A SPP is characterized by a phase term $\exp(il\varphi)$, where $\varphi$ is the azimuthal angular coordinate and $l$ is the azimuthal index or so-called the topological charge that can take any integer. This can be fabricated by grading the thickness of a dielectric layer deposited in the aperture proportional to the azimuthal angle around the centre of the plate, so that spiral phase shift sectors with $0 - 2\pi$ phase variation are achieved and repeated $l$ times, as shown in Fig. 1c ($l = +3$). In our design of a stepped SPP, the total phase shift of $2\pi$ is quantified into 8 levels in each phase shift sector. Fig. 1d pictures the SEM image of the etched SPP ($l = +3$), where the multi-sector, multi-step SPP structures are displayed. The SPPs were fabricated in a 1000 $nm$ thick silicon nitride ($n_r = 1.9$) film deposited on the top of the VCSEL and patterned using a Focused Ion Beam (FIB) etching technique. The total step height $h_s$ of the SPP is 826$nm$.

## Structures and Topological Charge Measurement Results

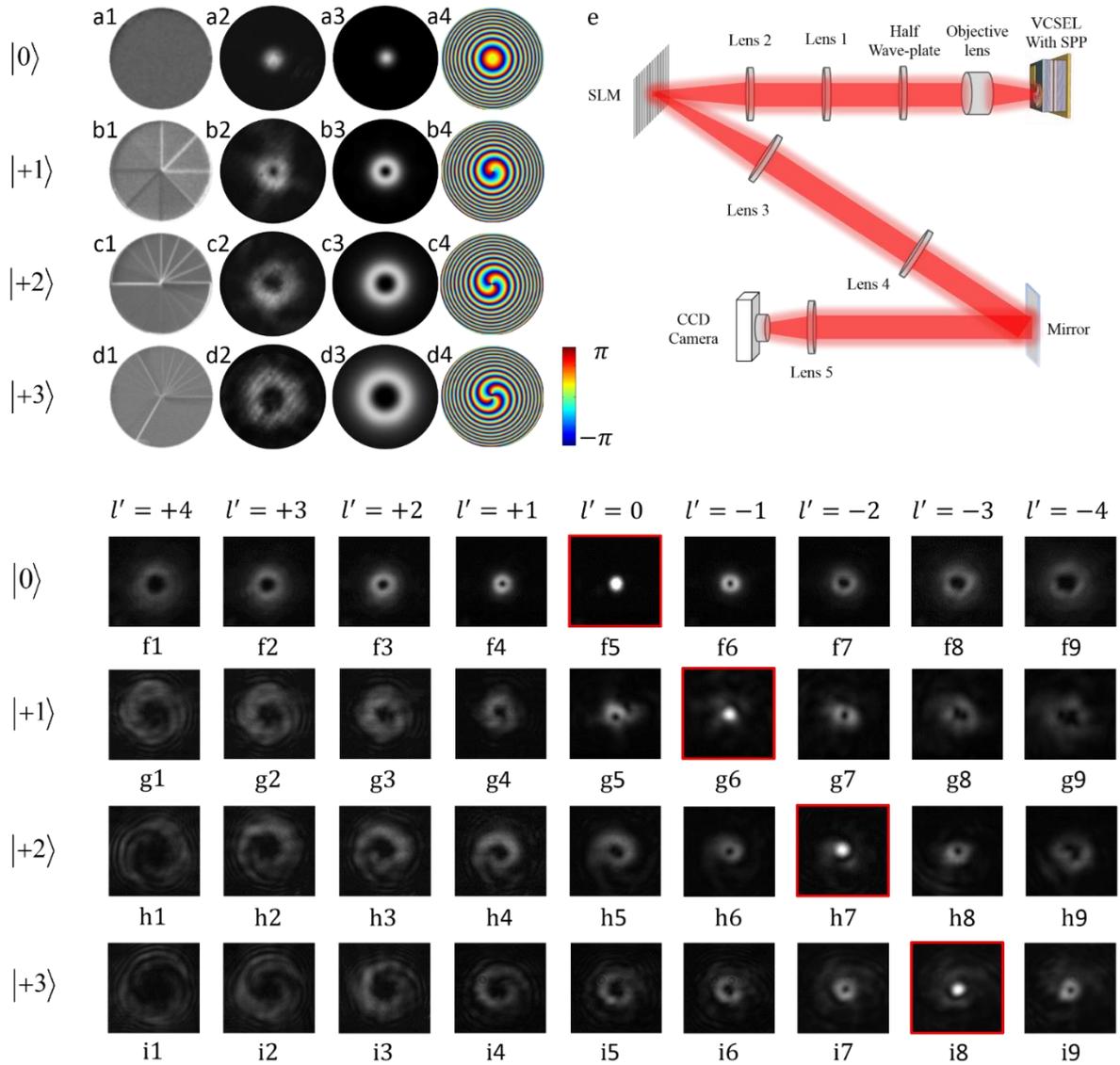

Figure 2. Experimental and simulation results of the OAM beams. (a1, b1, c1, d1): Top view SEM images of the fabricated SPP. (a2, b2, c2, d2): Experimentally observed intensity profiles of the generated OAM beams from the SPP. (a3, b3, c3, d3): Simulated intensity profiles of the OAM beams. (a4, b4, c4, d4): Simulated phase patterns of the OAM beams. (e): The experimental setup for characterization of OAM modes. (f, g, h, i): Intensity profiles of the OAM beam diffracted by the SLM.

The fabricated SPPs in the aperture of the VCSEL with various OAM mode orders are pictured in Fig. 2a1, b1, c1 and d1. The L-I and spectral characteristics of the VCSEL devices remain unchanged after the SPP fabrication process. To observe the farfield of the generated OAM beams, a modified 4f imaging system experimental setup is established[19], as shown in

Fig.2e. For $l = 0$, as shown in Fig. 2 row (a), the emitting area is only covered the silicon nitride film so the farfield of the generated beam remains Gaussian. For devices with SPP of single topological charge, well-defined farfield intensity patterns have been observed, as displayed in Fig. 2 b2, c2 and d2. These annular patterns are in very good agreement with those expected of optical vortex modes as given by semi-analytical simulations (Fig. 2 b3, c3 and d3). The higher the topological charge, the larger is the diameter of the annular farfield pattern. The dark spot on the optical axis can be attributed to the phase singularity. The simulated multi-level spiral phase patterns of the farfield have been illustrated in Fig. 2ab4, c4 and d4. It can be observed that the number of the helical arms is equal to the number $l$ of the helical phase front of the OAM beam.

The emitted beams are further characterised using computed holograms displayed on a spatial light modulator (SLM) to confirm their topological charge. The liquid-crystal-based SLM has dimensions of 7.68 × 7.68 mm, 512 × 512 pixels and a working wavelength range of 760nm-865nm. The analysis system can be separated into two subsystems, as shown in Fig 2e. The VCSEL emitter is placed at the front focal plane of the objective lens, which produces the Fourier transform of the emitted beam in its back focal plane. A lens (lens 1) is placed at a distance $f_1$, the same as its focal length, from the back focal plane of the objective. Another lens (lens 2) is placed in front of the SLM at a distance equaling the focal length of lens 2, projecting the farfield image of the VCSEL emission onto the hologram pattern. The inserted half-wave plate is used to rotate the polarization so that it matches that of the SLM hologram. After diffraction by the SLM, there is another modified 4f system (lens 3, lens 4 and lens 5) to project the diffracted beam on the CCD camera.

Fig.2 row (f), (g), (h) and (i) show the intensity profiles of the beams diffracted from the SLM, emitted by VCSELs integrated SPPs of $l=0$, +1, +2 and +3. The obtained images are new optical vortex beams with topological charge of $l' + l$, where $l'$ is the topological charge of the SLM hologram pattern. When an optical vortex beam of $l$ illuminates the SLM with an inverse spiral phase mask $l' = -l$, a Gaussian beam is produced shown as a bright spot in the centre. In the cases of $l' \neq l$, the farfield resumes an annular pattern. As an example, in Fig. 2 g6, the emitted beam through the SPP is with $l = +1$ and the SLM topological charge is set as $l' = -1$. A bright spot (Gaussian beam) appears at the centre of the image. This 'despiralization' of the beam can therefore confirm that the topological charge of the VCSEL-emitted beam is indeed $l = +1$, while the other ring-like patterns in row (g) are related to converted beams with topological charge of $l' + 1$. Similar results are shown for $l = 0$ (Fig. 2 row (f)), $l = +2$ (Fig. 2 row (h)) and for $l = +3$ (Fig. 2 row (i)).

As can be seen in Fig. 2 row (g), (h) and (i), some degradation of beam quality could be inferred. In theory, the SPP is not a pure OAM mode converter[20], and the stepped SPP further introduces purity degradation depending on the number of the steps in one phase sector[21]. The OAM mode purity of the generated beam has been studied by means of the SLM[3,22]. Figure 4 presents the simulated and measured mode purities for OAM beams generated by the VCSELs with different topological charges. The simulated OAM mode purity and weight values are calculated by the overlap-integral method[23,24]. In the case of $l = 0$, the measured purity of the OAM mode is about 95%, which could be attributed to factors such as limited flatness of the silicon nitride film, as well as measurement system error[3]. The mode purity values for $l = +1, l = +2,$ and $l = +3$ are down to 89%, 84%, and 78% compared with the simulation results of 98%, 97%, and 95%, respectively.

In the measured results, mode purity degradation could arise from symmetry-breaking factors, such as the less than ideal symmetry in the intensity distribution of the emitted beam from the VCSEL (i.e., deviation from Gaussian distribution), misalignment between the VCSEL emission and SPP's centers[3], and the imperfect uniformity of the fabrication. The degradation can also be attributed to the atmospheric turbulence as well as distortions in the free space optical systems [25] used to characterize the beams, which causes phase variation and hence the purity degradation of the OAM state. The purity of OAM modes is limited as a result of the generation of unwanted modes due to these non-ideal factors[26]. Device emission mode impurity could be improved by better device fabrication and would also require better optical characterization systems to be measured more precisely.

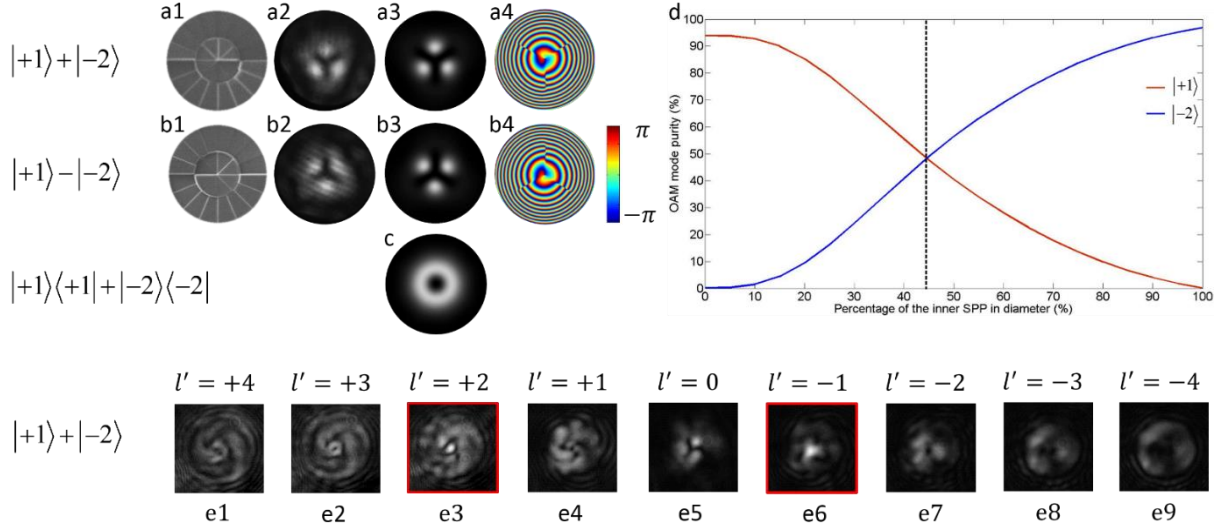

Figure 3. Experimental and simulation results of the OAM beams with superposition states. (a1, b1) Top view SEM images of the fabricated SPP. (a2, b2) Observed intensity profiles of the generated OAM beams from the SPPs. (a3, b3) Simulated intensity profiles of the OAM beams. (a4, b4) Simulated phase patterns of the OAM beams. (c) Simulated intensity profile of a mixed OAM state. (d) OAM mode purities with the ratio of diameter of the inner SPP. (e) The beam patterns of the OAM beam diffracted from the SLM.

In addition to realizing the single OAM states, a key strength of the approach is that it can generate an OAM superposition state comprising different topological charges. This is realized by designing a SPP with an inner circle ($|+1\rangle$) and an outer circle ($|-2\rangle$) that cover concentric regions of the VCSEL aperture, as shown in Fig. 3 a1 and b1. By setting the diameter of the inner circle to 44.6% (the black dashed line) of the full SPP's diameter (8.5$\mu m$), as shown in Fig. 3d, an equal weights of $|+1\rangle$ and $|-2\rangle$ components can be achieved as they each intercept an equal amount of light energy flux. By changing the relative rotation between the inner circle and the outer circle, the relative phase of $|+1\rangle$ and $|-2\rangle$ components can be tuned from 0 to $2\pi$.

Here we constructed two composite SPPs which can realize superpositions of $(|+1\rangle+|-2\rangle)$ and $(|+1\rangle-|-2\rangle)$, respectively, as shown in Fig. 3 row (a) and row (b). It can be seen in Fig 3 a2 and b2, the petal-like intensity patterns are confirmed with the theoretical predictions as shown in Fig.3 a3, a4, b3 and b4. These patterns are distinctively different from the simulated pattern of the mixed state of $|+1\rangle$ and $|-2\rangle$, as shown in Fig.3 c. This indicates that the SPP

generated beam is a coherent superposition of two different OAM components rather than a classical mixture of the two OAM lights.

To characterize the OAM components, again, we lead the generated beam passing through the OAM analysis system based on SLM. The characterization results of a superposition OAM state ($|+1\rangle+|-2\rangle$) are shown in Fig.3 row (e). A bright spot can be observed in two diffracted beam patterns (Fig.3 e3 and e6) corresponding to SLM topological charge of $l' = -1$ and $l' = +2$ indicating the simultaneous existence of the two OAM states. As show in Fig. 4e, the mode energy is nearly equally distributed into $|+1\rangle$ and $|-2\rangle$ with the values of 34% and 38% respectively, while the corresponding simulated values are 48% and 49%. The two experimental values add up to 82%, which is similar to that of 84% obtained with the single $l = +2$ mode.

The interference patterns of two OAM modes generated by conventional approaches have been calculated and demonstrated before[27–30], predicting fringe patterns with radial petals with ($l_{inner}$ -$l_{outer}$) number of lobes. However, the bulky and lossy optics inevitably lead to optical misalignments and therefore degrade the purity of the OAM states. From this point of view, this OAM VCSEL device is a promising candidate for OAM-based quantum communication systems. For example, [ $|+1\rangle$ , $|-2\rangle$ ] and [ ($|+1\rangle+|-2\rangle$) , ($|+1\rangle-|-2\rangle$) ], comprise mutually unbiased bases (MUBs) for a two dimensional QKD protocols and this scheme can straight forwardly upgrade to high-dimensional quantum systems[31,32].

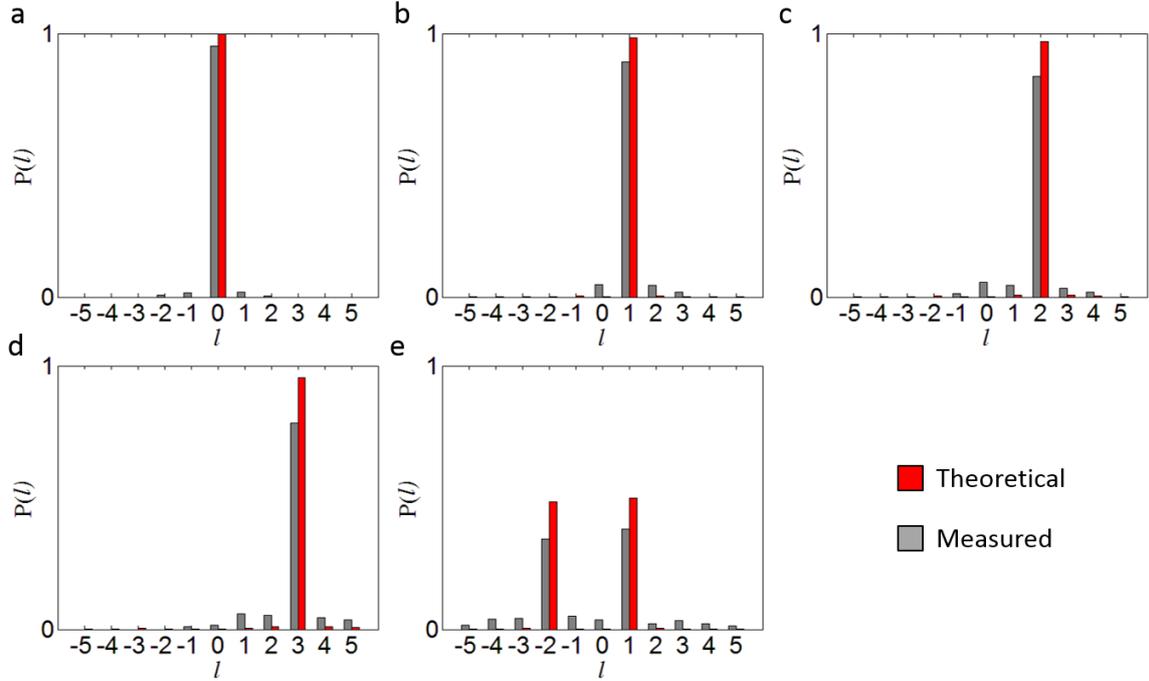

Figure 4. Mode purity calculated and measured results with (a) $l = 0$, (b) $l = +1$, (c) $l = +2$, (d) $l = +3$, (e) $l = |+1\rangle + |-2\rangle$.

## Conclusion

OAM lasers that emit beams with singular and superposition OAM states have been demonstrated based on single longitudinal and polarisation mode VCSEL devices integrated with micro-sized SPP. Although the generation of OAM beams using SPPs is a well-known technique, our innovative integrated approach enables low cost wafer scale fabrication that is especially attractive for array applications[8,9,24,33]. It is conceivable that large scale fabrication techniques, especially nano-imprinting, can be used to fabricated the integrated SPPs in polymeric materials in a single step, as the depth of all SPPs are the same regardless of their topological charge.

For many applications, the ability to emit multiple OAM modes in a concentric fashion is essential. In this regard, the scheme that can produce OAM superposition states is particularly attractive to QKD applications [17,18] using quantum superposition and transmitting information in OAM states[32]. In the future, more OAM states can be generated from one VCSEL device, making this scheme another alternative approaches to achieving a free-space quantum system[34,35]. Applications can also expand into other areas, such as optical trapping[36], quantum metrology[37], and remote detection[38], where the ability to maintain the coherence between the states is important.

The multi-OAM state emission scheme could be further extended by making electrically isolated concentric VCSEL apertures each with its own integrated SPP, in order to independently modulate the power of each of the many concentric OAM VCSELs with data. OAM multiplexed high capacity optical communications could thus be enabled with the reported modulation bandwidth of VCSEL extending well into several tens of gigahertz[39].